\documentclass[]{emulateapj}
\usepackage[dvips]{rotating}
\bibpunct{(}{)}{;}{a}{}{,}

\shorttitle{Deriving Spectroscopic Temperatures of Metal-Poor Stars}
\shortauthors{Frebel et al.}

\begin{document}
\title{Deriving Stellar Effective Temperatures of Metal-Poor Stars
  with the Excitation Potential Method\altaffilmark{1}}

\author{
Anna Frebel\altaffilmark{2},
Andrew R. Casey\altaffilmark{3,2},
Heather R. Jacobson\altaffilmark{2},
Qinsi Yu\altaffilmark{2}
}

\altaffiltext{1}{This paper includes data gathered with the 6.5 meter
  Magellan Telescopes located at Las Campanas Observatory, Chile.}

\altaffiltext{2}{Kavli Institute for Astrophysics and Space Research
  and Department of Physics, Massachusetts Institute of Technology, 77
  Massachusetts Avenue, Cambridge, MA 02139, USA}

\altaffiltext{3}{Research School of Astronomy \& Astrophysics, The
  Australian National University, Weston, ACT 2611, Australia}

\begin{abstract}

It is well established that stellar effective temperatures determined
from photometry and spectroscopy yield systematically different
results. We describe a new, simple method to correct spectroscopically
derived temperatures (``excitation temperatures'') of metal-poor stars
based on a literature sample with
$-3.3<\mbox{[Fe/H]}<-2.5$. Excitation temparatures were determined
from Fe\,I line abundances in high-resolution optical spectra in the
wavelength range of $\sim$3700 to $\sim$7000\,{\AA}, although shorter
wavelength ranges, up to 4750 to 6800\,{\AA}, can also be employed,
and compared with photometric literature temperatures. Our adjustment
scheme increases the temperatures up to several hundred degrees for
cool red giants, while leaving the near-main-sequence stars mostly
unchanged. Hence, it brings the excitation temperatures in good
agreement with photometrically derived values. The modified
temperature also influences other stellar parameters, as the
Fe\,I-Fe\,II ionization balance is simultaneously used to determine
the surface gravity, while also forcing no abundance trend on the
absorption line strengths to obtain the microturbulent velocity. As a
result of increasing the temperature, the often too low gravities and
too high microturbulent velocities in red giants become higher and
lower, respectively. Our adjustment scheme thus continues to build on
the advantage of deriving temperatures from spectroscopy alone,
independent of reddening, while at the same time producing stellar
chemical abundances that are more straightforwardly comparable to
studies based on photometrically derived temperatures. Hence, our
method may prove beneficial for comparing different studies in the
literature as well as the many high-resolution stellar spectroscopic
surveys that are or will be carried out in the next few years.

\end{abstract}

\section{Introduction}

Determining the atmospheric parameters of a star is fundamental to
characterizing and understanding its nature and evolutionary
state. These parameters are the effective temperature T$_{\rm eff}$,
surface gravity $\log$\,g, metallicity [M/H] and for 1D abundance
analyses also the microturbulent v$_{mic}$. Particularly, the
temperature can be determined with several methods, such as from
photometric colors, flux calibrated low-resolution spectra, the shape
of the Balmer lines in a high-resolution stellar spectrum and
through forcing no trend of Fe\,I absorption line abundances with the
excitation potential of the lines (``excitation balance''), also from
high-resolution spectra. The most common technique is to employ
broadband colors, with many color-temperature calibrations available
for main-sequence stars (e.g., \citealt{alonso_ms, casagrande10}) and
giants (e.g., \citealt{alonso_giants}) based on the infrared flux
method (IRFM). Precise photometry, reliable reddening values and
metallicity information are necessary when using this method. Using
flux-calibrated low-resolution spectra to measure the strength of the
Balmer jump and comparing the spectral shape to grids of synthetic
flux spectra is less common \citep{bessell07}.

In the absence of photometry or stellar libraries, or to test
systematic temperature uncertainties, temperatures can also be
obtained from high-resolution spectroscopy alone. Fitting the shape of
the Balmer lines with synthetic spectra of known stellar parameters
yields temperatures, especially for the warmer main-sequence stars and
subgiants (e.g., \citealt{barklem_balmer}). Finally, another commonly
used method is to employ Fe\,I lines throughout the stellar spectrum
and force their abundances to show no trend with the excitation
potential of each line. Compared to using photometry to derive
temperatures, no reddening information is needed for this technique or
the Balmer line fitting method. But in turn, high-resolution spectra
with resolving power of $R\gtrsim15,000$ are required, preferably with
large wavelength coverage throughout the optical range. A list of Fe
absorption lines covering a large range of excitation potentials is
also needed.

The different methods are known to produce temperatures with
systematic offsets from each other. The ``excitation temperatures''
are thereby known to yield lower effective temperatures than
photometry-based color-temperature calibrations, often by a few
hundred degrees. Stars on the upper red giant branch are most
affected. \citet{johnson2002_23stars} found their 23 metal-poor giants
to be 100 to 150\,K cooler when using the excitation method over
several color-temperature calibrations. Similar results have been
reported by \citet{cayrel2004}, \citet{aoki_studiesIV}
\citet{lai2008}, \citet{ufs}, \citet{hollek11}, and many others. The
approaches and dealings with this issue have been varied, though.

\citet{cayrel2004} reached good agreement between their
photometrically derived temperatures and excitation temperatures after
they excluded strong lines with excitation potential of
$EP=0$\,eV. These often appear to have higher abundances than lines with
higher excitation potential. As a consequence, forcing no trend in
line abundances with excitation potential can yield artificially
cooler temperatures. While a similar approach (cutting $EP<0.2$\,eV)
worked for \citet{cohen08}, it did not yield the same outcome for
\citet{lai2008} and \citet{hollek11}. With or without cutting low-$EP$
lines, the latter two studies found significant trends of abundance
with excitation potential to remain. Consequently, different authors
have adopted either spectroscopic temperatures, or the photometric
ones, or tried to find a middle ground between the results of the two
techniques.

The situation is further complicated by the fact that different model
atmosphere codes have been used over time, producing different results
leading to and possibly increasing the differences between spectroscopic
and photometric temperatures. For example, using a model code that
treated scattering as true absorption \citet{cayrel2004} found larger
discrepancies than when using codes that treated scattering as
Rayleigh scattering.

However, recently, \citet{hollek11} carried out an abundance analysis of
16 metal-poor giants using different versions of the MOOG code
\citep{moog}. They found that if scattering is properly treated, the
temperatures were generally even lower than in an older MOOG version that
treated scattering as true absorption -- the opposite of what was
found by \citet{cayrel2004} (although they make no quantitative
statement about the magnitude of the effect and how it compares to
their proper-scattering treatment when not excluding lines with
$EP<1.2$\,eV). 

This study aims at empirically resolving the issue of producing lower
spectroscopic excitation temperatures when employing the latest
version of the widely-used MOOG code that now properly treats the
scattering \citep{sobeck11}. We provide simple corrections to the
slope of the Fe\,I abundances as a function of excitation potential
for metal-poor stars that bring the excitation temperatures in rough
agreement with photometrically derived values. This way, we can use
``the best of both worlds'': not depending on photometry, reddening
and color-temperature relations, while obtaining temperatures and
hence chemical abundances that can be easily and relatively fairly
compared to results based on photometric temperatures. Our pragmatic
approach does not resolve the underlying processes that likely cause
the temperature discrepancies which seem to be manifold (scattering
treatments in codes, line formation of low and high-$EP$ lines,
wavelength coverage of lines and data quality, effects due to
non-local thermodynamic equilibrium, etc.) but at least it provides a
means to facilitate a more homogeneous analysis of stars that have
temperatures derived from different methods. Much work has been
invested in improving photometric temperature scales. For example,
using the IRFM method, \citet{casagrande10} showed this photometric
temperature determination method to be hardly sensitive to theoretical
model parameters, especially the surface gravity and metallicity
(particularly for extremely metal-poor stars). The uncertainty with
respect to their zero-point (well-calibrated, against interferometric
angular diameter measurements and HST spectroscopy) using solar twins
is 15\,K, whereas their final IRFM effective temperatures have typical
uncertainties ranging from 60 to 90\,K for metal-poor stars and
slightly lower values for more metal-rich stars. The independence of
metallicity when deriving temperatures this way makes the IRFM method,
in principle, the preferred one for low-metallicity stars. The only
major caveat would be that the IRFM method sensitively depends on very
accurate reddening values, as an 0.01\,mag change in $E(B-V)$ leads to
$\pm50$\,K change in the final temperature.

Given these facts, it appears timely to bring the somewhat plagued
excitation temperatures onto a comparable scale. Another reason for
such a ``re-calibration'' is that the stellar surface gravities,
usually derived through the Fe\,I-Fe\,II ionization equilibrium,
depend on the derived temperatures. With lower temperatures, lower
gravities are derived. Hence, in many cases of cool giants, much too
low gravities have been obtained (compared with isochrones and
parallax-derived values), to a point that they become unphysical. This
effect is particularly pronounced in metal-poor stars with lower $S/N$
spectra which leave few Fe\,II to be measured (e.g., see Table~5 in
\citet{hollek11}). In the same way, if only primarily strong lines are
measurable in the spectrum, the microturbulent velocity becomes very
large, often past 2.5\,km\,s$^{-1}$.  Especially, metal-poor stars in
dwarf galaxies have been affected by this, since they are all bright
giants and the data quality is often rather poor given their faint
apparent magnitudes \citep{mcwilliametal, koch_her, aoki09, hollek11,
  leo4}.

In summary, adjusting the spectroscopic temperatures leads to improved
stellar parameters, at least in the metal-poor regime and particularly
for metal-poor giants with $\mbox{[Fe/H]}<-2.0$. Broadly adopting this
scheme may prove beneficial for the many high-resolution
spectroscopic surveys (Gaia-ESO, GALAH) that will soon
provide large amounts of stellar spectra that have to be analysed
homogeneously and consistently.

\section{``Calibration'' Sample and Abundance Analysis}

We chose seven metal-poor stars with metallicities of
$-3.3<\mbox{[Fe/H]}<-2.5$ and spanning the largest possible temperature
range from 4600\, to 6500\,K covering warm main-sequence to cool
giants to investigate systematic differences between the excitation
temperatures and photometric temperatures.

All high $S/N$, high-resolution spectra were obtained with the MIKE
spectrograph \citep{mike} on the Magellan-Clay telescope at Las
Campanas Observatory between 2009 and 2011. An $0\farcs7$ slit width
was used resulting in a resolving power of $R\sim35,000$ in the blue
and $\sim28,000$ in the red spectral range. MIKE spectra cover nearly
the full optical wavelength range of 3350--9100\,{\AA}, although the
$S/N$ is low below 3700\,{\AA}. The reduced spectra were normalized,
the orders merged and then radial velocity corrected. The $S/N$ ratio
of the final spectra ranges from 70 to 300 at $\sim4100$\,{\AA}, and
from 70 to 500 at $\sim6700$\,{\AA} per pixel.

\begin{figure}[!hb]
 \begin{center}
   \includegraphics[clip=true,width=8cm,bbllx=30, bblly=360,
     bburx=452, bbury=724]{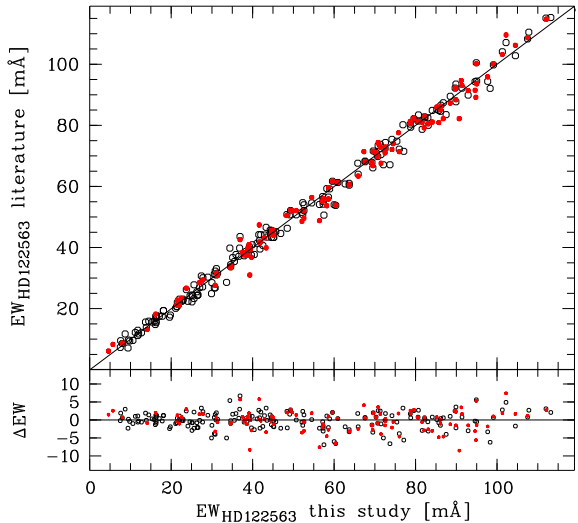} \figcaption{ \label{ew_comp}
     Comparison of equivalent width measurements for HD122563 of this
     study with those of \citet{aoki_studiesIV} (black open circles)
     and \citet{cayrel2004} (red filled circles). The agreement is excellent.}
 \end{center}
\end{figure}

Equivalent widths were measured by fitting Gaussian profiles to the
absorption lines. For a metal line list we used the list described in
\citet{roederer10}. Between $\sim$50 and $\sim$230 Fe\,I and 2 and 26
Fe\,II lines were measured in each sample star. See
Table~\ref{Tab:eqw} for the lines used and their measured equivalent
widths, although we only list lines with reduced equivalent widths of
$\gtrsim-4.5$. In Figure~\ref{ew_comp} we furthermore show a
comparison of the $\sim200$ equivalent width measurements of HD122563
common to our lines and those of \citet{aoki_studiesIV} and
\citet{cayrel2004}. Not all our lines were measured by these two
studies, but those that are in common agree very well with each
other. The agreements of our measurements are $0.20\pm0.16$\,m{\AA} and
$0.25\pm0.28$\,m{\AA}, respectively, so no significant offsets exist
between the three data sets.

\begin{deluxetable*}{llrrrrrrrrrrrrrrrrr}
%\rotate
\tablecolumns{11}
\tablewidth{0pt}
\tabletypesize{\tiny}
\tablecaption{\label{Tab:eqw} Equivalent Widths and Metal Line Abundances of Stars in the Calibration Sample}
\tablehead{
\colhead{} &
\colhead{} &
\colhead{} &
\colhead{} &
\multicolumn{2}{c}{HD122563} & 
\multicolumn{2}{c}{HE~1523$-$0901} &
\multicolumn{2}{c}{BD$-$18 5550} & 
\multicolumn{2}{c}{CS22892$-$052} &
\multicolumn{2}{c}{HD140283} & 
\multicolumn{2}{c}{CD$-$24 17504} &
\multicolumn{2}{c}{G64$-$12} 
\\ 
\colhead{$\lambda$} &
\colhead{Species} &
\colhead{$\chi$} &
\colhead{$\log{gf}$} &
\colhead{EW} &
\colhead{$\log\epsilon(\rm X)$} &
\colhead{EW} &
\colhead{$\log\epsilon(\rm X)$} &
\colhead{EW} &
\colhead{$\log\epsilon(\rm X)$} &
\colhead{EW} &
\colhead{$\log\epsilon(\rm X)$} &
\colhead{EW} &
\colhead{$\log\epsilon(\rm X)$} &
\colhead{EW} &
\colhead{$\log\epsilon(\rm X)$} &
\colhead{EW} &
\colhead{$\log\epsilon(\rm X)$} 
\\
\colhead{[{\AA}]} & 
\colhead{} & 
\colhead{[eV]} & 
\colhead{[dex]} &
\colhead{[m{\AA}]} & 
\colhead{[dex]} & 
\colhead{[m{\AA}]} & 
\colhead{[dex]} & 
\colhead{[m{\AA}]} & 
\colhead{[dex]} & 
\colhead{[m{\AA}]} & 
\colhead{[dex]} & 
\colhead{[m{\AA}]} &
\colhead{[dex]} & 
\colhead{[m{\AA}]} & 
\colhead{[dex]} & 
\colhead{[m{\AA}]} & 
\colhead{[dex]} }
\startdata
 5889.950&   Na\,I&  0.00&   0.108& 196.64&   4.00& 175.91&   3.54& 141.05&   3.52& 153.99&   3.72& 112.99&   3.84&  29.91&   2.70&  35.99&   2.94 \\
 5895.924&   Na\,I&  0.00&$-$0.194& 166.54&   3.89& 155.12&   3.51& 117.04&   3.37& 126.36&   3.53&  89.59&   3.70&  19.75&   2.76&  18.90&   2.86 \\
 3829.355&   Mg\,I&  2.71&$-$0.208&\nodata&\nodata&\nodata&\nodata&\nodata&\nodata&\nodata&\nodata& 132.91&   5.17&  85.21&   4.68&  77.74&   4.55 \\
 3986.753&   Mg\,I&  4.35&$-$1.030&  33.86&   5.22&\nodata&\nodata&  19.58&   4.99&\nodata&\nodata&  15.29&   5.26&\nodata&\nodata&\nodata&\nodata \\
...
\enddata
\end{deluxetable*}

For our 1D LTE abundance analysis, we used model atmospheres from
\citet{castelli_kurucz} and the MOOG analysis code of \citet{moog},
albeit the latest version that appropriately treats scattering as
Rayleigh scattering and not as true absorption as done in previous
versions \citep{sobeck11}. The metal line abundances are presented in
Table~\ref{Tab:eqw}. Our study about stellar parameter determinations
is based on equivalent width measurements of Fe lines between
3700 and 7000\,{\AA}.

\section{Adjusting the ``Excitation Temperature'' Scale}\label{sec:adjust}

We used our sample of well-studied stars to produce a straight-forward
method of adjusting the excitation temperature scale in order to arrive
at spectroscopic temperatures that agree with photometric ones. After
measuring the equivalent widths we first determined the stellar
temperatures by forcing no trend of the abundances of individual Fe\,I
lines as a function of their excitation potential.  In the process we
varied the microturbulent velocity to produce no trend of the line
abundances with reduced equivalent widths. Simultaneously, to
determine the surface gravity, the Fe\,II abundance was matched to the
Fe\,I abundance. Finally, another input parameter of the model
atmosphere is the metallicity [m/H] of the star. We use a general
prescription of $\mbox{[m/H]} = \mbox{[Fe/H]} +0.25$ to account for the
overabundances of $\alpha$-elements of $\mbox{[$\alpha$/Fe]}\sim0.3$
as well as small overabundances of carbon and/or oxygen
($\mbox{[C,O/Fe]}\sim0.3$) which are all typical for metal-poor halo
stars. 

We then carried out an extensive literature search to assess the range
of photometric temperatures determined for each star.
Table~\ref{Tab:teff_lit} shows a representative range of values found
in the literature, including the colors and the color-temperature
relations used for their calculation.  For comparison, we also include
T$_{\rm eff}$ values calculated using the IRFM method in recent works
whose color- temperature relations are commonly used (e.g., Alonso et
al.\ 1996, 1999, Casagrande et al.\ 2010).  For two stars, CD$-$24
17504 and G64$-$12, we found relatively few photometrically derived
values, and most where done using older temperature calibrations. We
thus calculated photometric temperatures ourselves, based on
literature photometry and reddening values and using the
\citet{casagrande10} scale. Finally, for HE~1523$-$0901, only one
photometric temperature is available \citep{he1523} but at least it
was an average based on six different colors.

As can been seen, temperature ranges of several hundred degrees were
found for some cases, making it difficult to assess which individual
temperatures would be ``the best''.  Hence, we resorted to choosing
``common sense'' values, as our original aim was to re-calibrate our
excitation temperature scale to a level common with that of
photometric temperatures found using the more recent color-temperature
relations. Our initial excitation temperatures and associated adopted
literature values (based on photometry) are listed in
Table~\ref{Tab:stell_par_calib}.  We stress that the photometric
temperatures we have adopted are not strictly averages of the values
listed in Table~\ref{Tab:teff_lit}, rather they are appropriate
near-median quantities from the range of values found by multiple
authors using different color-temperature relations and photometry.

\begin{figure}[!hb]
 \begin{center}
   \includegraphics[clip=true,width=8cm,bbllx=30, bblly=260,
     bburx=462, bbury=784]{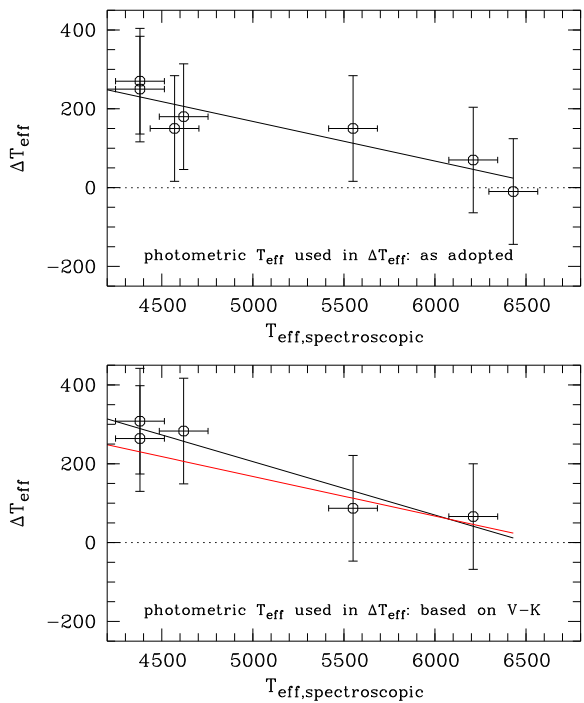}
   \figcaption{ \label{calib} Differences in effective temperatures
     between our initial excitation temperatures and adopted
     literature photometry-based values (top panel) and those
     calculated from $V-K$ (bottom panel). We also show the slope from
     the top panel in the bottom panel (red line) The differences are
     most pronounced at the cooler temperatures of upper red giant
     branch stars, whereas main-sequence stars are hardly affected.}
 \end{center}
\end{figure}

Figure~\ref{calib} (upper panel) then shows the differences in
effective temperatures between our initial excitation temperatures and
our adopted literature values as a function of the initial
temperatures. We find a strong, linear relation, with the largest
differences at the cooler temperatures of upper red giant branch
stars. An (unweighted) least-square regression yields the following
relation:

\begin{equation}
T_{\rm{eff, corrected}} = T_{\rm{eff, initial}} - 0.1 \times T_{\rm{eff, initial}} + 670
\end{equation}

This temperature dependent behavior is completely in line with the
vast majority of previous studies that reported differences between
spectroscopic and photometric temperatures of several hundred degrees,
as most analyzed star were giants. However, main-sequence stars are
hardly affected, as we find essentially the same temperatures. We also
overplot nominal error bars of 134\,K which reflect typical
uncertainties in temperature determinations (60\,K for photometric and
120\,K for spectroscopic values).

\begin{deluxetable*}{llllll}
\tablecolumns{6}
\tablewidth{0pt}
\tabletypesize{\tiny}
\tablecaption{\label{Tab:teff_lit} Effective Temperatures Collected from Literature}
\tablehead{
\colhead{$T_{\rm{eff}}$ } &
\colhead{$\sigma$} &
\colhead{Colors} &
\colhead{Method} &
\colhead{Reference} &
\colhead{Comments}} 
\startdata  
\multicolumn{6}{c}{HD122563}\\\hline
 4670 & 28\tablenotemark{1}& V,J,K       & COH78, FRO 83     & \citet{peterson90}     & phot. from NIC78, CAR83\\
 4650 & 30\tablenotemark{2}& B,V,R,I     & BG78, BG89        & \citet{ryan96}         & their phot., BN84, GS91\\
 4572 & 61                 & IRFM        & their calibration & \citet{alonso_giants}  & as quoted in PASTEL database\\
 4617 & 44\tablenotemark{1}& B,V,R,J,K   & ALO99             & \citet{cayrel2004}     & phot. from ALO98, BEE07, EPC99; adopt 4600\\
 4546 & 75\tablenotemark{1}& B,V,K       & ALO99             & \citet{honda04}        & phot. from Simbad; adopt 4570\\
 4616 & 17\tablenotemark{1}& B,V,K,b,y   & ALO99             & \citet{lai}            & phot. from Simbad, HM98 \\
 4843 & \nodata            & B,V,R,J,K   & RM05              & \citet{yong13_II}      & phot. from CAY04, HON04\\\hline\\
\multicolumn{6}{c}{HE~1523$-$0901}\\\hline
 4630 & 40                 & B,V,R,I,J,H,K& ALO99            & \citet{he1523}         & phot. from BEE07\\\hline\\
\multicolumn{6}{c}{BD$-$18 5550}\\\hline
 4750 &  0\tablenotemark{1}& V,J,K       & COH78, FRO83      & \citet{peterson90}     & phot. from CAR78, CAR83\\ 
 4789 & 27\tablenotemark{1}& B,V,R       & BG78, BG89        & \citet{mcwilliametal}  & their phot.\\
 4668 & \nodata            & IRFM        & their calibration & \citet{alonso_giants}  & as quoted in PASTEL database \\
 4700 & 129\tablenotemark{1}&B,V,R,J,K   & ALO99             & \citet{cayrel2004}     & phot. from ALO98, BEE07, EPC99; adopt 4750\\
 4558 & \nodata            & B,V,R,J,K   & RM05              & \citet{yong13_II}      & phot. from CAY04\\\hline\\
\multicolumn{6}{c}{CS~22892$-$052}\\\hline
 4763 & 54\tablenotemark{1}& B,V,R       & BG78, BG89        & \citet{mcwilliametal}  & their phot.\\
 4850 & \nodata            & B,V         & see RYA96b        &\citet{1997norriscarbon}& phot. from RYA89, BPS92, NOR97\\
 4860 & 83\tablenotemark{1}& B,V,R,I,J,K & ALO99             & \citet{cayrel2004}     & phot. from  ALO98, BEE07, EPC99; adopt 4850\\
 4740 & 69\tablenotemark{1}& B,V,K       & ALO99             & \citet{honda04}        & adopt 4790\\
 4825 & \nodata            & B,V,R,J,K   & RM05              & \citet{yong13_II}      & phot. from CAY04, HON04\\\hline\\
\multicolumn{6}{c}{HD140283}\\\hline
 5691 & \nodata            & IRFM        & their calibration & \citet{alonso_ms}      & as quoted in HOS09 \\
 5640 & \nodata            & V,R,I,K     & their calibration & \citet{bikmaev96}      & as quoted in MAS99 \\
 5750 & 50\tablenotemark{2}& B,V,R,I     & BO86, BK92        & \citet{ryan96}         & their phot.\\
 5750 & 53                 & V,J,K       & HOU00             & \citet{cohen02}        & their phot.\\
 5609 & 34\tablenotemark{1}& B,V,K       & ALO96             & \citet{honda04}        & phot. from Simbad; adopt 5630\\
 5751 & 50\tablenotemark{2}& uvby-$\beta$& ALO96             & \citet{jonsell05}      & phot. from OLS83, SN88\\
 5777 & 55                 & IRFM        & their calibration & \citet{casagrande10}   & \\
 5711 & \nodata            &             & CAS10             & \citet{yong13_II}      & phot. from COH02, HON04\\\hline\\
\multicolumn{6}{c}{CD$-$24 17504}\\\hline
 6060 & 54\tablenotemark{2}& B,V,R,I,b,y & BO86, BK92, MAG87,& \citet{ryan96a}        & phot. from RYA89\\
      &                    &             & VB85              &                        &  \\
 6373 & \nodata            & IRFM        & their calibration & \citet{alonso_ms}      & as quoted in PRI00b\\
 6237 & 79\tablenotemark{1,3}& b,y       & CAR83, KIN93      & \citet{primas}         & phot. from RYA99 \\
 6455 & 62                 & IRFM        & their calibration & \citet{casagrande10}   & \\ 
 6236 & \nodata            &             & CAS10             & \citet{yong13_II}      & phot. from RYA91\\
 6290 & \nodata            & V,R,I,J,K   & CAS10             & This Study             & phot. from RYA91\\\hline\\\\
\multicolumn{6}{c}{G64$-$12}\\ \hline
 6325 & \nodata            & B,V,R,I,b,y &BUS87, BO86, KUR89,& \citet{ryanSDIV91}     & phot. from RYA89, SN88, CAR83, this work \\
      &                    &             & MAG87, VB85       &                        &  \\
 6290 & 44\tablenotemark{2}& B,V,R,I,b,y & BO86, BK92, MAG87,& \citet{ryan96a}        & phot. from CAR78, CAR89, CS92, SN89, SCH93\\
      &                    &             & VB85              &                        &  \\
 6470 & 90                 & IRFM        & their calibration & \citet{alonso_ms}      & as quoted in PRI00a\\
 6430 & 75\tablenotemark{1}& B,V,R,I,K   & ALO96             & \citet{Aokihe1327}     &    \\
 6464 & 61                 &             & IRFM              & their calibration      & CAS10 \\
 6450 & 35\tablenotemark{1}& B,V,R,I,K   & CAS10             & This Study             & phot. from AOK06 colors \\
\enddata
\tablenotetext{1}{Standard deviations of the average temperatures (column 1) calculated from the different colors (column 3).}
\tablenotetext{2}{Error based on assessment of errors in photometry/reddening.}
\tablenotetext{3}{Average of values from two different color-T$_{\rm eff}$ relation.}

\tablecomments{JHK magnitudes were used in many of the cited studies and most are taken from 2MASS \citep{2MASS}.
References for photometry and color-temperature relations: 
ALO96 -- \citet{alonso_ms}; 
ALO98 -- \citet{alonso98};
ALO99 -- \citet{alonso_giants}; 
AOK06 -- \citet{Aokihe1327}; 
BO86  -- \citet{bell86};
BEE07 -- \citet{beers_photom};
BG78  -- \citet{bell78}; 
BG89  -- \citet{bell89}; 
BK92  -- \citet{buser92};
BN84  -- \citet{cd38};
BPS92 -- \citet{BPSII}; 
BUS87 -- Buser, private communication in \citet{ryanSDIV91};
BK92  -- \citet{buser92}; 
CAR78 -- \citet{carney78}; 
CAR83 -- \citet{carney83}; 
CAR89 -- \citet{carney89}; 
CAS10 -- \citet{casagrande10}; 
CAY04 -- \citet{cayrel2004};
CS92  -- \citet{cayreldestrobel92}; 
COH78 -- \citet{cohen78}; 
COH02 -- \citet{cohen02}; 
EPC99 -- \citep{epchtein99};
FRO83 -- \citet{frogel83}; 
GS91  -- \citet{gratton91};
HM98  -- \citet{hauck98};
HOU00 -- \citet{houdashelt00};
HON04 -- \citet{honda04};
HOS09 -- \citet{hosford09};
KIN93 -- \citet{king93}; 
KUR89 -- Kurucz, private communication in \citet{ryanSDIV91};
MAG87 -- \citet{magain87}; 
MAS99 -- \citet{mashonkina_ba_nlte};
NIC78 -- \citet{nicolet78};
NOR97 -- \citet{1997norriscarbon};
OLS83 -- \citet{olsen83};
PASTEL -- \citet{soubiran10};
PRI00a -- \citet{primas2000};
PRI00b -- \citet{primas};
RM05  -- \citet{ramirez05};
RYA89 -- \citet{ryan89}; 
RYA91 -- \citet{ryanSDIV91}; 
RYA96b -- \citet{ryan96}; 
RYA99 -- \citet{ryan_postprim}; 
SN88  --\citet{schuster88}; 
SN89  -- \citet{schuster89}; 
SCH93 -- \citet{schuster93}; 
VB85  -- \citet{vandenberg85}
}
\end{deluxetable*}

\begin{deluxetable*}{lcc|cccc|cccc}
\tablecolumns{7}
\tablewidth{0pt}
\tabletypesize{\small}
\tablecaption{\label{Tab:stell_par_calib} Stellar Parameters of the Sample}
\tablehead{
\colhead{}&
\colhead{}&
\colhead{}&
\multicolumn{4}{c}{initial determination}&
\multicolumn{4}{c}{after temperature correction}\\
\cline{4-7}\cline{8-11}
\colhead{Star}  &
\colhead{$T_{\rm{eff}, adop.}$ } &
\colhead{$T_{\rm{eff},V-K}$} &
\colhead{$T_{\rm{eff}}$ } &
\colhead{$\log (g)$ }    &
\colhead{$\mbox{[Fe/H]}$ }  &
\colhead{$v_{\rm{micr}}$ }  &
\colhead{$T_{\rm{eff}}$}&
\colhead{$\log (g)$ }    &
\colhead{$\mbox{[Fe/H]}$ }  &
\colhead{$v_{\rm{micr}}$ }\\
\colhead{}&
\colhead{[K]}&
\colhead{[K]}&
\colhead{[K]}&
\colhead{[dex]}&\colhead{[dex]}&\colhead{[km\,s$^{-1}$]}&
\colhead{[K]}&\colhead{[dex]}&\colhead{[dex]}&\colhead{[km\,s$^{-1}$]}
}
\startdata                           % init. stellpar  corrected stellpar                
 HD122563       & 4650 &  4644 & 4380 & 0.10 & $-2.93$& 2.65 & 4612& 0.85 & $-2.79$& 2.30 \\  
 HE~1523$-$0901 & 4630 &  4688 & 4380 & 0.05 & $-2.97$& 2.85 & 4612& 0.80 & $-2.85$& 2.70 \\  
 BD $-$18 5550  & 4720 &\nodata& 4570 & 0.80 & $-3.20$& 2.05 & 4783& 1.25 & $-3.02$& 2.00 \\  
 CS22892-052    & 4800 &  4903 & 4620 & 0.85 & $-3.24$& 2.20 & 4828& 1.35 & $-3.08$& 2.15 \\  
 HD140283       & 5700 &  5637 & 5550 & 2.95 & $-2.74$& 1.50 & 5665& 3.20 & $-2.64$& 1.45 \\  
 CD $-$24 17504 & 6280 &  6276 & 6210 & 3.60 & $-3.26$& 1.35 & 6259& 3.65 & $-3.23$& 1.40 \\  
 G64$-$12       & 6420 &\nodata& 6430 & 4.35 & $-3.30$& 1.45 & 6457& 4.40 & $-3.28$& 1.50     
\enddata
%\tablecomments{}
\end{deluxetable*}

Since we adopted common sense representative photometric
temperatures from the literature that are based on a variety of colors
and color-temperature relations, we also attempted to calculate
temperatures ourselves, for an exemplary color, $V-K$, and using the
\citet{alonso_giants} and \citet{alonso_ms} calibrations. We chose
$V-K$ since it is least affected by reddening. The reddening in the
direction of each object was obtained from the dust maps of
\citet{schlegel} maps. However, it is known that these maps
overpredict the reddening, in some cases quite
significantly. Accordingly, different reddening correction were
invoked, following the procedures of \citet{reddening_bonifacio} and
\citet{melendez08}. Temperatures were then calculated based on the
two reddening estimates to investigate the impact of these reddening
uncertainties. In a few cases, we adopted yet a different value.  For
HD140283, we also adopt $E(B-V)=0.01$, following \citet{ryan96},
rather than $E(B-V)=0.16$ from the dust maps. For HD122563, we also
adopt $E(B-V)=0.0$, following \citet{honda04}.

We also find a strong relation between the difference in temperatures
as a function of the initial spectroscopic temperatures, as can be
seen in Figure~\ref{calib} (bottom panel). The slope is even steeper
because $V-K$ produces temperatures for giants that are warmer than
what we adopted. Within the error bars, however, the slope of our main
temperature comparison is still in rough agreement with what we
determined based on $V-K$ colors. We note that this comparison only
contains five of the seven sample stars. We had to exclude two
stars (BD$-$18 5550 and G64$-$12) because their $V-K$ temperature were
rather different (more than 200\,K in the case of BD$-$18 5550) from
what one would expect the temperatures of these stars to be,
regardless of any employed method.

In the end we adopt the slope of our initial comparison because our
overall aim is to adjust the spectroscopic temperatures to
\textit{average} photometrical values. Even with photometry and
color-temperature relations available, the temperatures derived from
different colors often vary 100-200\,K, as can be seen in
Table~\ref{Tab:teff_lit}. 
Finding a different slope based on the $V-K$ temperatures appears to
reflect this dispersion.

\begin{figure}[!hb]
 \begin{center}
 \includegraphics[clip=true,width=7.5cm,bbllx=40, bblly=100,
   bburx=502, bbury=710]{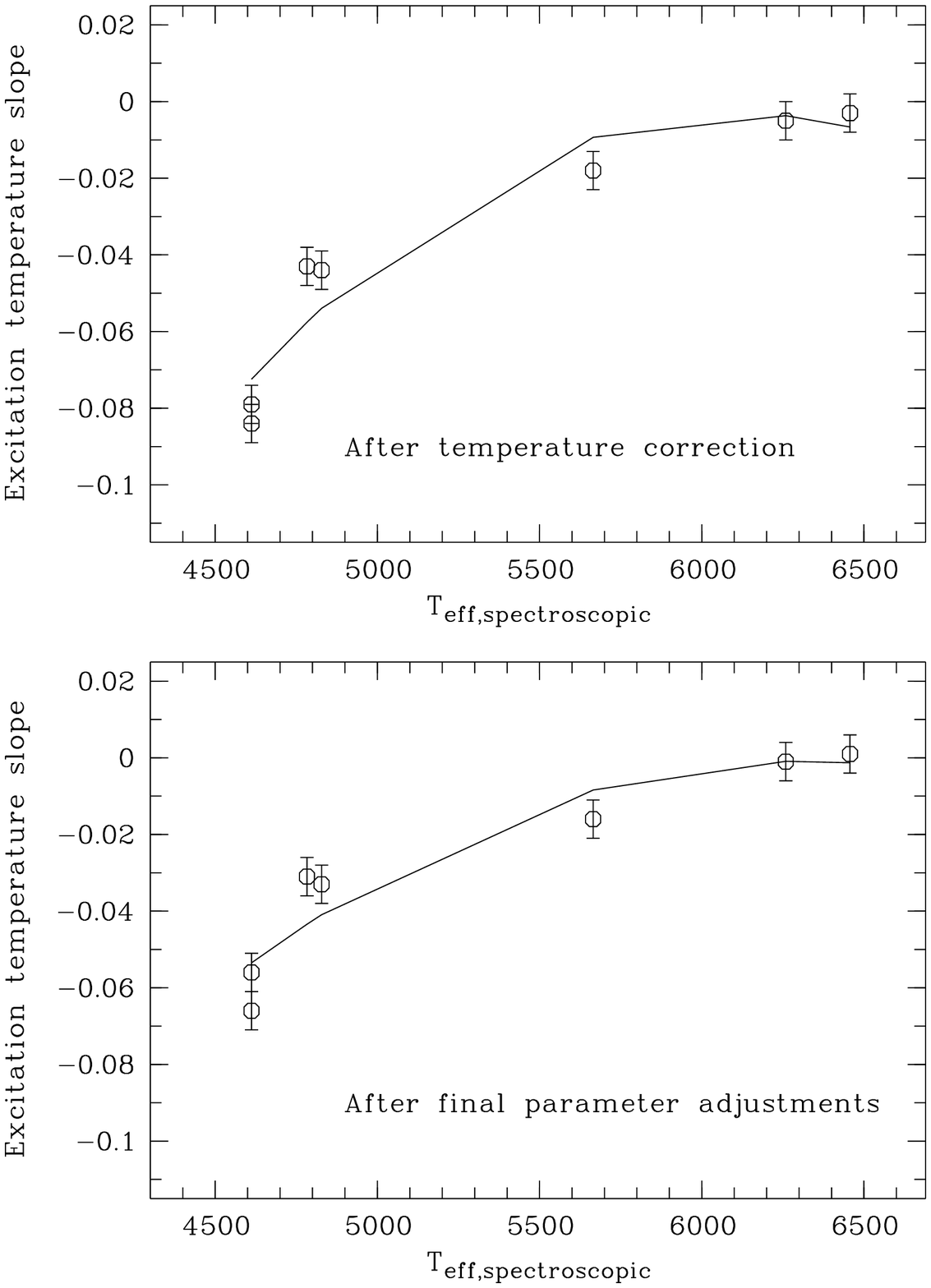} \figcaption{ \label{slope}
   The slopes of the Fe\,I line abundances as a function of their
   excitation potentials after only adjusting the temperature
   according to our method (top panel) and after also adjusting the
   surface gravity and microturbulent velocity (bottom panel) to
   arrive at the final stellar parameters. The maximum slope in the
   bottom panel nominally corresponds to $\sim150$\,K compared with a
   slope of 0.000\,dex/eV.}
 \end{center}
\end{figure}

We then turned the temperature differences from Figure~\ref{calib}
into values for the slope of the abundance trends to adopt in the
Fe\,I line abundance vs. excitation potential diagnostics. This is
shown in Figure~\ref{slope}. In the top panel we show how the values
of the slopes of the Fe\,I line abundances after only correcting for
the temperature and before adjusting the microturbulent velocity and
the gravity. To guide the eye, we fit the data points with a
second-order polynomial, which is also shown in the Figure. We
estimate typical uncertainties in the slope to be 0.005\,dex/eV or
less.

For near main-sequence stars, the spectroscopic method delivers nearly
the same temperatures as photometry, hence there is no significant
slope of line abundances with excitation potential
($\sim0.00$\,dex/eV). For the giants, however, applying significant
temperature corrections according to Eq. (1) produces an initial slope
of up to $-$0.09\,dex/eV.  We then carried out the remaining parts of
the analysis: re-determining the microturbulent velocity and the
surface gravity while ignoring the slope in the temperature diagnostic
plot of line abundance vs. excitation potential. While the gravity was
adjusted by up to $\sim0.7$\,dex for the most extreme cases, the
microturbulent velocity decreased by less ($\sim$0.1 or less in most
cases). The model metallicity [m/H] was usually increased by about 0.1
to 0.2\,dex following the increase in the Fe abundance.

Through the process of adjusting the remaining stellar parameters
(gravity, microturbulent velocity, metallicity), we actually found
that the slope of abundance with excitation potential was
reduced. Figure~\ref{slope} (bottom panel) shows the final slopes in
the temperature diagnostic plots. The most extreme one is now only
$-$0.07\,dex/eV which, in the end, only corresponds to a temperature
difference of 150\,K compared to a slope of 0.000\,dex/eV. However, we
caution that if one would change the temperature again to produce no
trend, the gravity and microturbulent velocity would change again, and
one would once again iterate towards the lower temperature/lower
gravity/higher microturbulent velocity solution. Regardless, 150\,K is
of the same order as what uncertainties in spectroscopic effective
temperatures generally are. It is thus reassuring that despite of
applying our temperature corrections, we are producing temperatures
that nominally are no more than 150\,K away from a trend with no
slope. Table~\ref{Tab:stell_par_calib} shows the initial and corrected
final sets of stellar parameters of all sample stars.

\begin{figure*}
 \begin{center}
   \includegraphics[clip=true,width=16cm,bbllx=30, bblly=165,
     bburx=552, bbury=774]{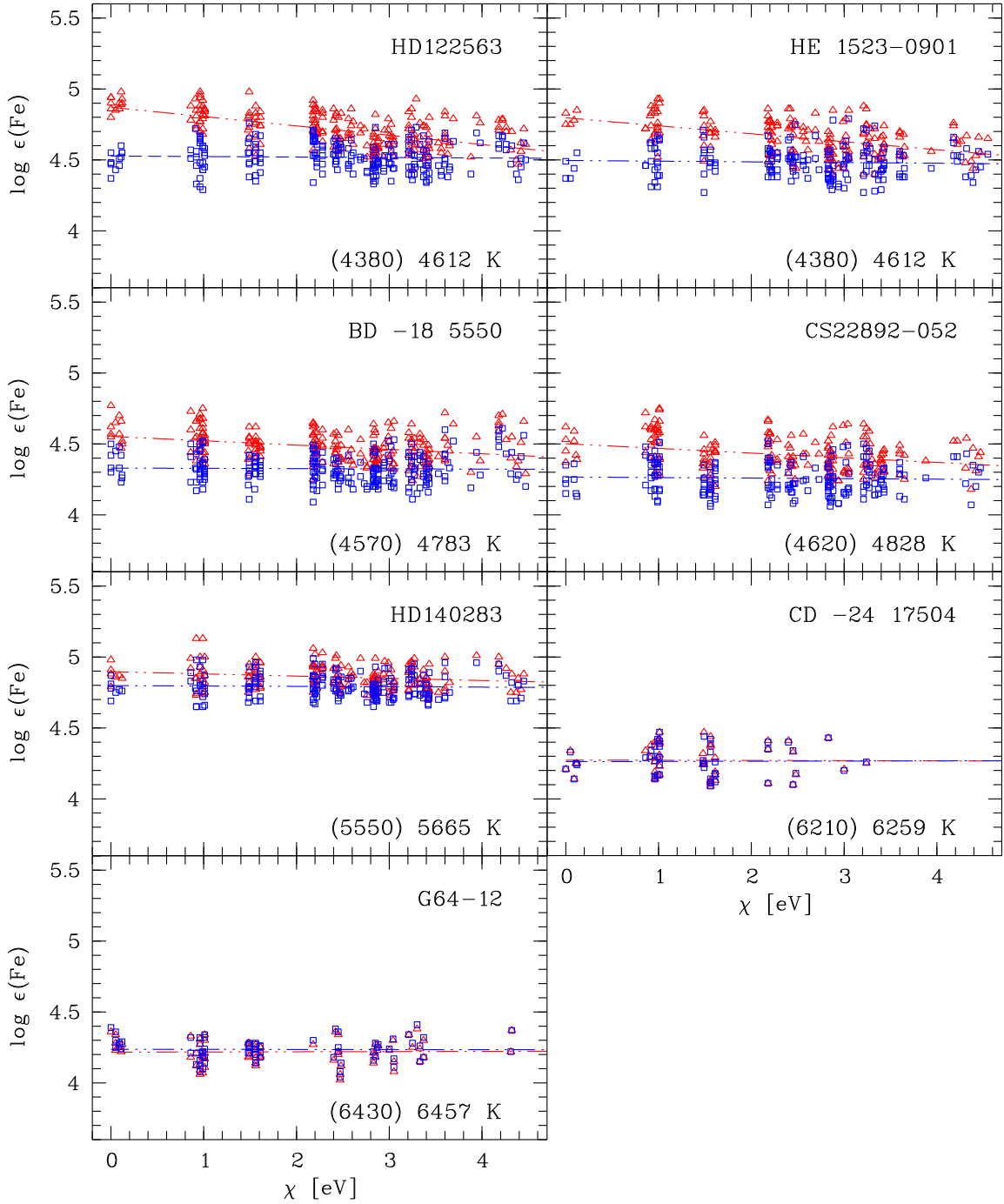} \figcaption{ \label{texcit}
     Fe\,I line abundances as a function of excitation potential
     $\chi$ for all sample stars. Plotted are the line abundances
     showing no trend of excitation potential as initially derived
     (blue squares), and the final line abundance trends following the
     temperature adjustment and after v$_{mic}$, $\log$\,g and
     metallicity have been adjusted (red triangles). Shown also are
     the corresponding initial (in brackets) and final temperature.}
 \end{center}
\end{figure*}

In Figure~\ref{texcit}, we show the Fe\,I line abundances as a
function of excitation potential for all sample stars. Plotted are the
line abundances showing no trend of excitation potential as initially
derived and the final line abundances after temperature correction and
subsequent adjustment of v$_{mic}$ and $\log$\,g. This illustrates how
the slope changes as a result of our analysis, which is also
quantified in the two panels in Figure~\ref{slope}.

\begin{figure}[!ht]
 \begin{center}
 \includegraphics[clip=true,width=8cm,bbllx=70, bblly=460, bburx=512,
   bbury=770]{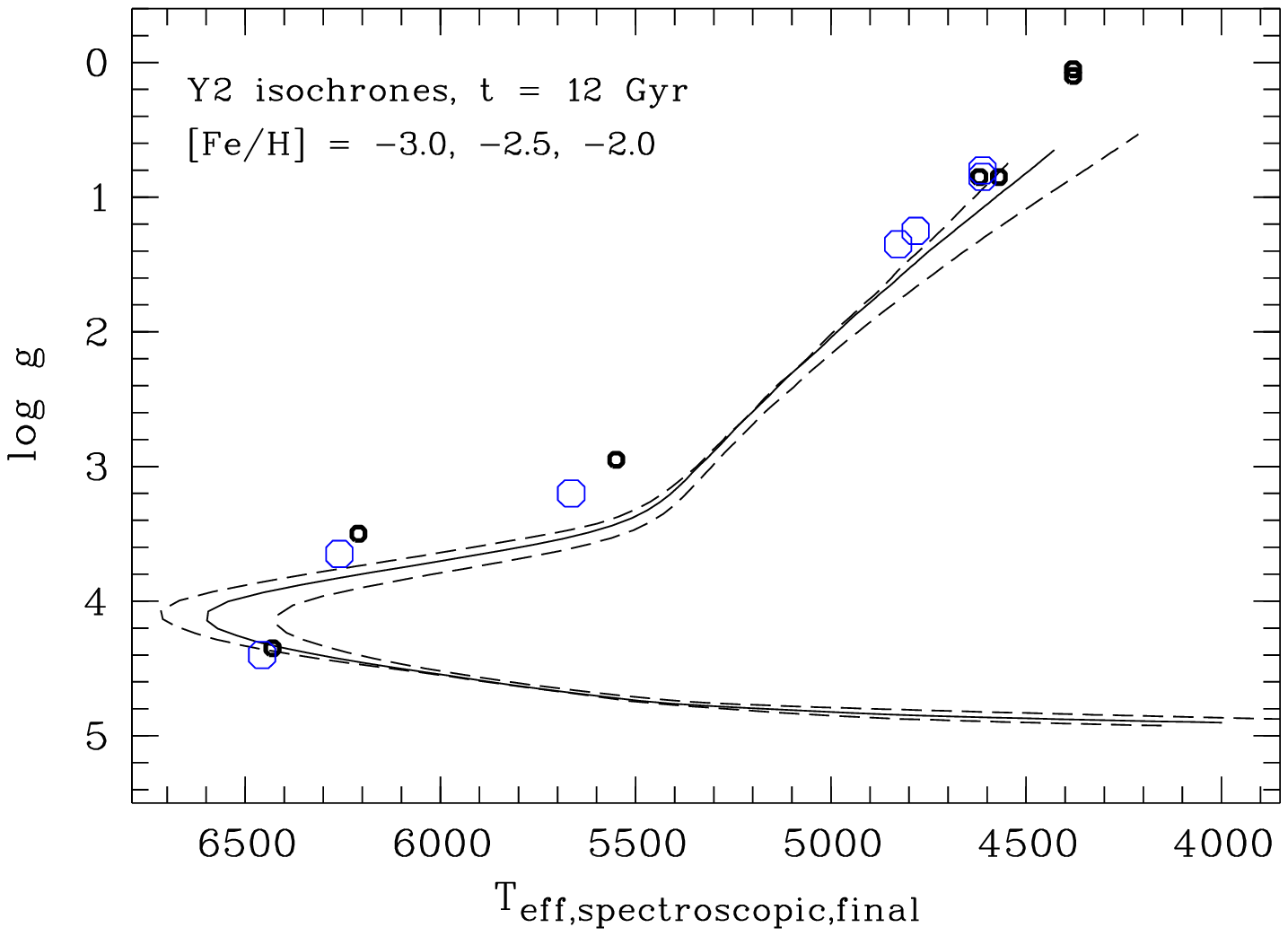} \figcaption{ \label{calib_iso}
   Our initial (small black open circles) and corrected stellar
   parameters (larger open blue circles) plotted over 12\,Gyr
   isochrones with $\mbox{[Fe/H]}=-3.0$, $\mbox{[Fe/H]}=-2.5$ and
   $\mbox{[Fe/H]}=-2.0$, all having an $\alpha$-enhancement of
   $\mbox{[$\alpha$/Fe]}=0.3$ \citep{Y2_iso}. The agreement of the
   stellar parameters with the isochrones following the temperature
   adjustment is very good. The cooler giants do not extend past
   the upper red giant branch of the isochrones anymore.}
 \end{center}
\end{figure}

We then checked how the initial and corrected sets of stellar
parameters compare with an isochrone. We used 12\,Gyr isochrones with
$\mbox{[Fe/H]}=-3.0$, $-2.5$ and $-2.0$ and $\alpha$-enhancement of
$\mbox{[$\alpha$/Fe]}=0.3$ \citep{Y2_iso}. Figure~\ref{calib_iso}
shows the results. The initial, too cool temperatures pushed the
giants off the isochrones at $<4500$\,K. The corrected values agree
very well with the isochrones. The only small exception is the
subgiant HD140283, which is $\sim$0.3\,dex off the subgiant branch or
250\,K from the base of the giant branch. However, the initial stellar
parameters made this star more of a warm giant rather than a
subgiant. Hence, the corrected stellar parameters provide better
solution for this star.

\begin{figure}[!ht]
 \begin{center}
   \includegraphics[clip=true,width=8cm,bbllx=50, bblly=430,
     bburx=472, bbury=724]{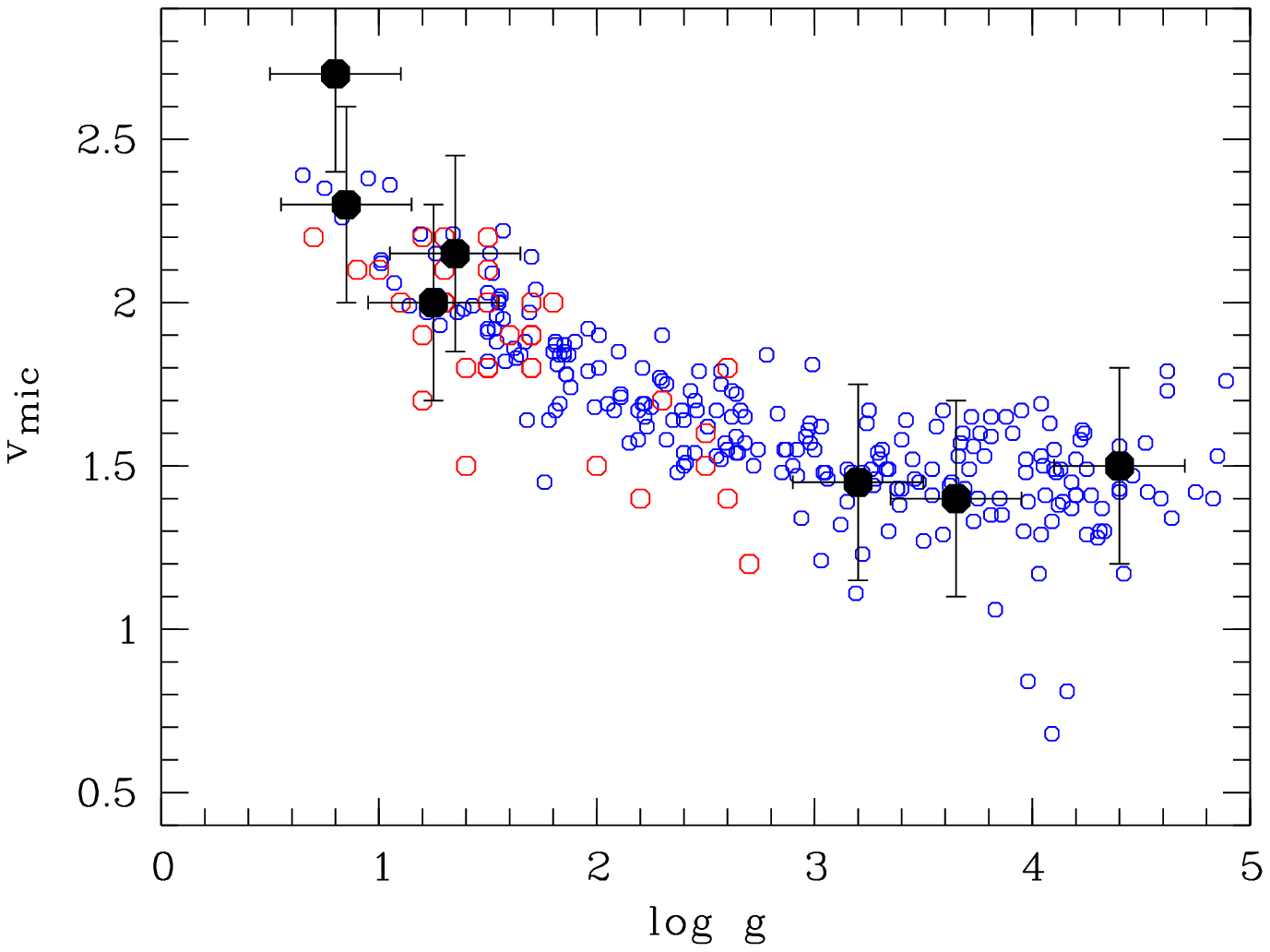} \figcaption{ \label{vmic}
     Final microturbulent velocities as function of the surface
     gravity for the sample stars (black filled circles) as well as
     the \citet{cayrel2004} sample (small red circles) and
     \citet{heresII} sample (without the horizontal branch stars;
     small blue circles).}
 \end{center}
\end{figure}

Finally, we checked how the microturbulent velocities as a function of
the surface gravity compare with those of other studies. In
Figure~\ref{vmic} we compare our values with those of
\citet{cayrel2004} and \citet{heresII} who used photometrically
determined temperatures for their studies. There exists some scatter
in the microturbulent velocities for a given gravity within each study
and perhaps also some small systematic difference. While the Cayrel et
al. values have been derived with the same methodology as used in this
study, the Barklem et al. values are derived from synthesizing
spectral lines by simultaneously solving for all four stellar
parameters (although based on appropriate initial guesses). Table~3 of
\citet{heresII} suggests that with this technique slightly higher
surface gravities and higher microturbulent velocities (0.1 to
0.3\,km\,s$^{-1}$) are derived, compared to other literature studies.
Assuming typical uncertainties of 0.3\,km\,s$^{-1}$ our microturbulent
velocities are in good agreement with those from the literature.

Based on the properties of our chosen sample of stars, this method has
been accurate for metal-poor subgiants and giants with
$-3.3<\mbox{[Fe/H]}<-2.5$ that have high-resolution spectra with a
large wavelength coverage (at least 4750 to 6800\,{\AA}). 

We carried out further tests to asses the validity of our adjustment
scheme at $\mbox{[Fe/H]}<-3.5$. As part of this we analyzed CD$-$38
245 ($\mbox{[Fe/H]}=-4.2$; e.g., \citealt{yong13_II}), HE~1300+0157
($\mbox{[Fe/H]}=-3.7$; \citealt{frebel_he1300}), HE~0557$-$4840
($\mbox{[Fe/H]}=-4.8$; \citealt{he0557}), HE~0107$-$5240
($\mbox{[Fe/H]}=-5.4$; \citealt{HE0107_Nature}), and HE~1327$-$2326
($\mbox{[Fe/H]}=-5.6$; \citealt{HE1327_Nature}). In all cases, we
found that the excitation temperatures yield much lower values
($\Delta$T$_{\rm eff}$ of 500 to 1400\,K) for these ultra metal-poor
stars than what was found for our sample of seven extremely metal-poor
stars (the maximal $\Delta$T$_{\rm eff}$ is 270\,K, see
Figure~\ref{calib}). Hence, we speculate that at $\mbox{[Fe/H]}<-3.5$
the excitation method to determine effective temperatures may not be
sufficient for the task, perhaps because of (metallicity dependent)
effects, such as NLTE effects or the accuracy of model atmospheres
(e.g. plane-parallel models for cooler, extended giants), or simply
because at those metallicities, only relatively few Fe lines are
available from the spectra.  While we used all known stars with
$\mbox{[Fe/H]}<-4.0$ that had published equivalent widths, the number
of stars remains small, and our individual $\Delta$T$_{\rm eff}$
values should perhaps be regarded as somewhat anecdotal. In summary,
we suggest that our calibration appears to be breaking down at
$\mbox{[Fe/H]}\sim-4.0$.

We also tested if any of our results would change if fewer blue lines
would be available due to the use of spectrographs other than MIKE on
Magellan or Xshooter on VLT that yield (nearly) full optical
coverage. Mimicking a Subaru/HDS two-arm setting (4000 to 6800\,{\AA})
and the VLT/UVES 580nm setting (4750 too 6800\,{\AA}), we checked
whether using reduced wavelength ranges would change our stellar
parameters. Only small changes were found. Temperatures changed by up
to 40\,K (making the stars warmer, although not in all cases) and the
microturbulent velocity changed between 0.05 and
0.3\,km\,s$^{-1}$. The main concern with this exercises was to have
sufficient number of Fe\,I lines (between 50 and 150) and Fe\,II lines
(more than at least 2 to 3) available, many of which (especially
Fe\,II) are located between 4400 and 4650\,{\AA}. Hence, for the near
main-sequence turn-off stars cutting at 4750\,{\AA} did not yield
well-determined stellar parameters since only 20 (for G64-12) and 6
(for CD$-$24 17504) Fe\,I and only 2 Fe\,II (both stars) were
available.

Finally, we note that any line abundance corrections based on line
formation under non-local thermodynamic equilibrium (NLTE) would
increase the surface gravity. We made a test where we assumed a simple
average correction for Fe\,I of 0.2\,dex, and assuming that NLTE does
not affect Fe\,II (e.g., \citealt{asplund_araa}). This increases the gravity of about
0.5\,dex.  Such a large change might be welcome for the cooler giants
when not applying our adjustments, i.e. in combination with the too
cool temperatures that result in too low gravities, but for
near-main-sequence stars a large gravity increase would pose problems
regardless. More detailed stellar parameter determination techniques
seem to be required to resolve this issue (e.g. \citealt{bergemann12,
  lind12}).

\section{Metal Abundance Results and Comparison with Literature Values}\label{ab_pa}

For completeness, we present the chemical abundances of our sample
stars in Table~\ref{abund}. Solar abundances are taken from
\citet{asplund09}. We note that while for the Fe abundances we employed
only lines with wavelengths longer than 3700\,{\AA}, but for a few
other elements (Ti, Cr, Ni) we also included a few lines at shorter
wavelengths for better statistics whenever they were measurable. For
each star, we also compared the abundances with those of another study
from the literature \citep{Norrisetal:2001, honda04,
  cayrel2004,Aokihe1327}. We chose those studies because they yielded
stellar parameters for a given star that agreed well with our
results. We took the abundances from \citet{frebel10} to ensure that
all studies employ the same solar abundances. The stellar abundance
differences in the final [X/Fe] values between our and the literature
studies are reasonable and typically within 0.2\,dex.

 \begin{deluxetable*}{lrrrrrrrrrrrrrrrrrrrrrrrrrrrrr}
% \rotate
 \tabletypesize{\tiny}
 \tablewidth{0pc}
\tablecaption{\label{abund} Magellan/MIKE Chemical Abundances of the Sample Stars}
 \tablehead{
 \colhead{}&
 \multicolumn{4}{c}{HD122563}&\colhead{}& \multicolumn{4}{c}{HE~1523$-$0901}&\colhead{}& \multicolumn{4}{c}{BD $-$18 5550}&\colhead{}& \multicolumn{4}{c}{CS22892-052}\\
 \cline{2-5} \cline{7-10}\cline{12-15}\cline{17-20}
 \colhead{Species} &
 \colhead{$\lg\epsilon (\mbox{X})$} & \colhead{[X/Fe]}& \colhead{$N$}& \colhead{$\sigma$}&\colhead{}&
 \colhead{$\lg\epsilon (\mbox{X})$} & \colhead{[X/Fe]}&  \colhead{$N$}& \colhead{$\sigma$}&\colhead{}&
 \colhead{$\lg\epsilon (\mbox{X})$} & \colhead{[X/Fe]}&  \colhead{$N$}& \colhead{$\sigma$}&\colhead{}&
 \colhead{$\lg\epsilon (\mbox{X})$} & \colhead{[X/Fe]}&  \colhead{$N$}& \colhead{$\sigma$} }
 \startdata
  Na\,I & 3.95 &     0.50 &     2 &0.05 &&  3.52 &   0.13 &  2 & 0.02 &&   3.45 &   0.23 &   2 & 0.07 &&   3.62 &    0.46&   2 &   0.10  \\
  Mg\,I & 5.26 &     0.45 &     7 &0.07 &&  5.08 &   0.33 &  6 & 0.09 &&   4.99 &   0.41 &   7 & 0.06 &&   4.85 &    0.33&   5 &   0.09  \\
  Si\,I & 5.17 &     0.45 &     3 &0.06 &&  5.09 &   0.43 &  1 & 0.00 &&   4.90 &   0.41 &   1 & 0.00 &&   5.05 &    0.62&   1 &   0.00  \\
  Ca\,I & 3.87 &     0.32 &    25 &0.08 &&  3.78 &   0.29 & 21 & 0.08 &&   3.72 &   0.40 &  25 & 0.08 &&   3.66 &    0.39&  18 &   0.08  \\
  Sc\,I & 0.42 &     0.07 &    13 &0.07 &&  0.27 &$-$0.03 & 12 & 0.08 &&   0.17 &   0.04 &   9 & 0.06 &&   0.03 & $-$0.04&   6 &   0.07  \\
  Ti\,I & 2.28 &     0.12 &    31 &0.07 &&  2.26 &   0.16 & 26 & 0.09 &&   2.15 &   0.22 &  28 & 0.09 &&   2.07 &    0.19&  13 &   0.09  \\
  Ti\,II& 2.37 &     0.21 &    51 &0.10 &&  2.35 &   0.24 & 41 & 0.11 &&   2.16 &   0.23 &  49 & 0.11 &&   2.10 &    0.22&  36 &   0.10  \\
  Cr\,I & 2.56 &  $-$0.29 &    21 &0.08 &&  2.65 &$-$0.15 & 14 & 0.11 &&   2.43 &$-$0.19 &  20 & 0.03 &&   2.26 & $-$0.30&  10 &   0.06  \\
  Cr\,II& 2.89 &     0.04 &     3 &0.05 &&  3.12 &   0.32 &  2 & 0.12 &&   2.73 &   0.11 &   3 & 0.05 &&   2.45 & $-$0.11&   1 &   0.00  \\
  Fe\,I & 4.71 &     0.00 &   222 &0.12 &&  4.65 &   0.00 &183 & 0.12 &&   4.48 &   0.00 & 233 & 0.11 &&   4.42 &    0.00& 191 &   0.12  \\
  Fe\,II& 4.72 &     0.01 &    26 &0.10 &&  4.65 &$-$0.00 & 25 & 0.13 &&   4.49 &   0.01 &  24 & 0.08 &&   4.44 &    0.01&  21 &   0.10  \\
  Co\,I & 2.42 &     0.22 &     6 &0.10 &&  2.49 &   0.34 &  4 & 0.20 &&   2.24 &   0.27 &   8 & 0.10 &&   2.20 &    0.29&   7 &   0.10  \\
  Ni\,I & 3.56 &     0.13 &    24 &0.08 &&  3.40 &   0.03 & 17 & 0.09 &&   3.35 &   0.15 &  22 & 0.11 &&   3.36 &    0.22&  11 &   0.09  \\
  Zn\,I & 1.92 &     0.15 &     2 &0.04 &&  1.92 &   0.20 &  2 & 0.04 &&   1.86 &   0.33 &   2 & 0.05 &&   2.06 &    0.58&   1 &   0.00  \\\hline
 \colhead{}&
 \multicolumn{4}{c}{HD140283}&\colhead{}& \multicolumn{4}{c}{CD$-$24 17504}&\colhead{}& \multicolumn{4}{c}{G64$-$12}\\
 \cline{2-5} \cline{7-10}\cline{12-15}
 \colhead{Species} &
 \colhead{$\lg\epsilon (\mbox{X})$} & \colhead{[X/Fe]}& \colhead{$N$}& \colhead{$\sigma$}&\colhead{}&
 \colhead{$\lg\epsilon (\mbox{X})$} & \colhead{[X/Fe]}&  \colhead{$N$}& \colhead{$\sigma$}&\colhead{}&
 \colhead{$\lg\epsilon (\mbox{X})$} & \colhead{[X/Fe]}&  \colhead{$N$}& \colhead{$\sigma$} \\\hline
  Na\,I &    3.77 &     0.17  &   2 &   0.07 &&    2.73 &  $-$0.28  &  2 &    0.03 &&    2.90  & $-$0.06 &   2 &    0.04 \\
  Mg\,I &    5.26 &     0.30  &   9 &   0.07 &&    4.75 &     0.38  &  5 &    0.08 &&    4.70  &    0.38 &   6 &    0.09 \\
  Si\,I &    5.41 &     0.54  &   2 &   0.03 &&    4.40 &     0.12  &  1 &    0.00 &&    4.52  &    0.29 &   1 &    0.00 \\
  Ca\,I &    4.00 &     0.30  &  20 &   0.07 &&    3.32 &     0.21  &  3 &    0.06 &&    3.58  &    0.52 &  13 &    0.08 \\
  Sc\,I &    0.54 &     0.03  &   7 &   0.05 &&    0.21 &     0.29  &  1 &    0.00 &&    0.17  &    0.30 &   1 &    0.00 \\
  Ti\,I &    2.55 &     0.25  &  20 &   0.08 &&    3.12 &     1.40  &  2 &    0.02 && \nodata  & \nodata &   0 & \nodata \\
  Ti\,II&    2.52 &     0.21  &  44 &   0.08 &&    1.85 &     0.13  &  9 &    0.06 &&    2.18  &    0.51 &  14 &    0.07 \\
  Cr\,I &    2.83 &  $-$0.17  &  14 &   0.07 &&    2.26 &  $-$0.15  &  5 &    0.06 &&    2.22  & $-$0.14 &   6 &    0.08 \\
  Cr\,II&    3.09 &     0.09  &   3 &   0.08 &&    2.71 &     0.30  &  1 &    0.00 && \nodata  & \nodata &   0 & \nodata \\
  Fe\,I &    4.86 &     0.00  & 185 &   0.08 &&    4.27 &     0.00  & 49 &    0.11 &&    4.22  &    0.00 &  70 &    0.08 \\
  Fe\,II&    4.87 &     0.01  &  19 &   0.07 &&    4.29 &     0.01  &  2 &    0.02 &&    4.24  &    0.02 &   3 &    0.01 \\
  Co\,I &    2.59 &     0.24  &   6 &   0.05 &&    2.48 &     0.72  &  3 &    0.05 &&    2.17  &    0.46 &   2 &    0.02 \\
  Ni\,I &    3.77 &     0.20  &  22 &   0.08 &&    3.33 &     0.33  &  7 &    0.09 &&    3.19  &    0.25 &  10 &    0.09 \\
  Zn\,I &    2.05 &     0.14  &   2 &   0.05 && \nodata &  \nodata  &  0 & \nodata && \nodata  & \nodata &   0 & \nodata \\
 \enddata
 \tablecomments{ [X/Fe] abundance ratios are computed with [Fe\,I/H]
   abundances of the respective stars. Solar abundances have been
   taken from \citet{asplund09}. The uncertainties listed are standard
   deviations of the line abundances for each element. Standard errors
   for each abundance will be much smaller. For abundances measured
   from only one line, we adopt a nominal uncertainty of 0.10\,dex.}
\end{deluxetable*}

\begin{figure}[!hb]
 \begin{center}
 \includegraphics[clip=true,width=8cm,bbllx=70, bblly=210, bburx=512,
   bbury=770]{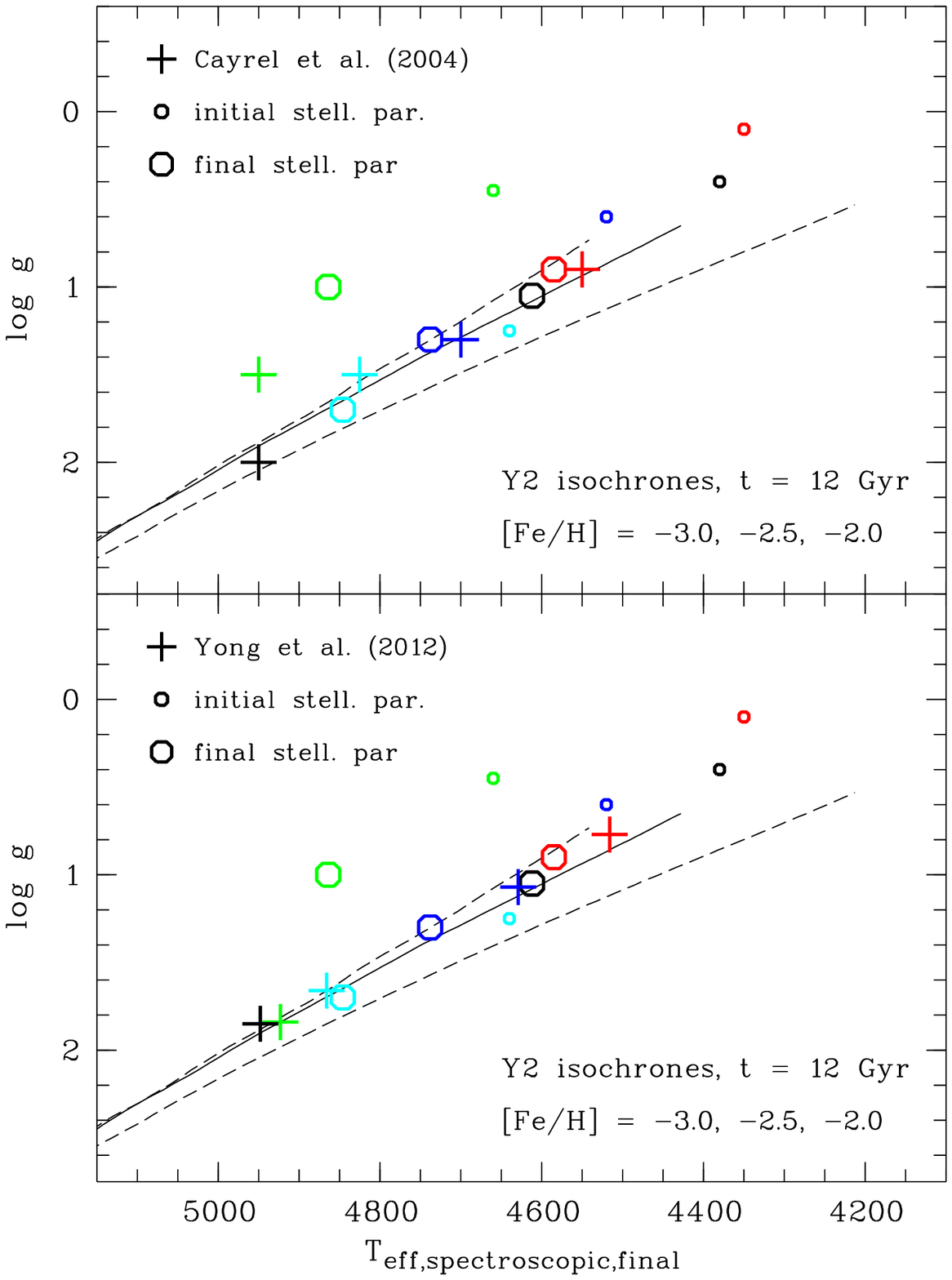}
 \figcaption{ \label{iso_literature} Initial (small open circles),
   corrected (larger open circles) stellar parameters compared with
   literature values (crosses) of \citet{cayrel2004} (top panel) and
   \citet{yong13_II} (bottom panel). Color code: green: HD2796, red:
   CS22873-166, black: CS30325-094,  blue: HD186478,
   cyan: CS31082-001. Overplotted are 12\,Gyr isochrones with
   $\mbox{[Fe/H]}=-3.0$, $\mbox{[Fe/H]}=-2.5$ and $\mbox{[Fe/H]}=-2.0$
   (from left to right), all having an $\alpha$-enhancement of
   $\mbox{[$\alpha$/Fe]}=0.3$ \citep{Y2_iso}. See text for
   discussion. }
 \end{center}
\end{figure}

Table~\ref{Tab:stell_par_literature} lists stellar parameters for an
additional five stars from the sample of \citet{cayrel2004} for which
we also had Magellan/MIKE spectra. We used these stars to further test
our method and stellar parameter determination technique. Furthermore,
we add three of our sample stars (HD122563, BD $-$18 5550,
CS22892-052) because they are also part of the Cayrel et
al. sample. This allows up to compare our stellar parameters with the
Cayrel et al. results as well as with those of \citet{yong13_II} who
analyzed the Cayrel et al. stars as part of a sample of 190
metal-poor stars from the literature. Figure~\ref{iso_literature}
illustrates the results and shows the good agreement of our effective
temperature and surface gravity compared to the photometrically
derived temperatures and gravities (also derived from the Fe\,I-Fe\,II
ionization equilibrium) by \citet{cayrel2004} (top panel) and
\citet{yong13_II} (bottom panel).

\begin{deluxetable}{lcccc} 
\tablecolumns{5}
\tablewidth{0pt}
\tabletypesize{\tiny}
\tablecaption{\label{Tab:stell_par_literature} Stellar Parameters of a Test Sample}
\tablehead{
\colhead{Star}  &
\colhead{$T_{\rm{eff}}$}&\colhead{$\log (g)$ }    &\colhead{$\mbox{[Fe/H]}$ }  &\colhead{$v_{\rm{micr}}$ }\\
\colhead{}&
\colhead{[K]} &\colhead{[dex]}&\colhead{[dex]}&\colhead{[km\,s$^{-1}$]}
}
\startdata   
\multicolumn{5}{c}{initial determination}\\\hline
%name         teff_   logg_ feh_o vmic_  
CS22873-166 &  4350 &   0.10 & $-$3.02 & 2.70  \\
HD122563    &  4380 &   0.10 & $-$2.93 & 2.65  \\
CS30325-094 &  4380 &   0.40 & $-$3.90 & 2.40  \\
HD186478    &  4520 &   0.60 & $-$2.64 & 2.10  \\
CD$-$38 245 &  4500 &   0.85 & $-$4.28 & 2.05  \\ 
BD$-$18 5550&  4570 &   0.80 & $-$3.20 & 2.05  \\
CS31082-001 &  4640 &   1.25 & $-$3.00 & 2.25  \\
CS22892-052 &  4620 &   0.85 & $-$3.24 & 2.20  \\
HD2796      &  4660 &   0.45 & $-$2.65 & 2.35  \\\hline

\multicolumn{5}{c}{after temperature correction}\\\hline
CS22873-166 &  4585 & 0.90 & $-$2.90 &2.60  \\
HD122563    &  4612 & 0.85 & $-$2.79 &2.30  \\
CS30325-094 &  4612 & 1.05 & $-$3.70 &2.15  \\
HD186478    &  4738 & 1.30 & $-$2.50 &2.15  \\
CD$-$38 245 &  4720 & 1.40 & $-$4.06 &1.95  \\
BD$-$18 5550&  4783 & 1.25 & $-$3.02 &2.00  \\
CS31082-001 &  4846 & 1.70 & $-$2.82 &2.25  \\
CS22892-052 &  4828 & 1.35 & $-$3.08 &2.15  \\
HD2796      &  4864 & 1.00 & $-$2.50 &2.25  \\\hline

\multicolumn{5}{c}{Cayrel et al. (2004)}\\\hline
CS22873-166 &  4550 & 0.90 &$-$2.97 & 2.10 \\
HD122563    &  4600 & 1.10 &$-$2.82 & 2.10 \\
CS30325-094 &  4950 & 1.50 &$-$3.30 & 2.00 \\
HD186478    &  4700 & 1.30 &$-$2.59 & 2.00 \\
CD$-$38 245 &  4800 & 1.50 &$-$4.19 & 2.20 \\
BD$-$18 5550&  4750 & 1.40 &$-$3.06 & 1.80 \\
CS31082-001 &  4825 & 1.80 &$-$2.91 & 1.50 \\
CS22892-052 &  4850 & 1.60 &$-$3.03 & 1.90 \\
HD2796      &  4950 & 2.10 &$-$2.47 & 1.50 \\\hline

\multicolumn{5}{c}{Yong et al. (2013)}\\\hline
CS22873-166 &  4516 & 0.77 &$-$2.74 & 1.60 \\
HD122563    &  4843 & 1.62 &$-$2.54 & 1.80 \\
CS30325-094 &  4948 & 1.85 &$-$3.35 & 1.60 \\
HD186478    &  4629 & 1.07 &$-$2.68 & 2.00 \\
CD$-$38 245 &  4857 & 1.54 &$-$4.15 & 2.20 \\
BD$-$18 5550&  4558 & 0.81 &$-$3.20 & 1.70 \\
CS31082-001 &  4866 & 1.66 &$-$2.75 & 1.40 \\
CS22892-052 &  4825 & 1.54 &$-$3.03 & 1.60 \\
HD2796      &  4923 & 1.84 &$-$2.31 & 1.60 
 \enddata 
\tablecomments{The original abundance from \citet{cayrel2004} have been
  redetermined using the solar abundances from \citet{asplund09}. }
\end{deluxetable}

The only star for which we determine very different stellar parameters
than \citet{cayrel2004} and \citet{yong13_II} is CS30325-094. We find
its corrected temperature to be about 340\,K cooler. A simple check of
the shape of the Balmer lines in comparison with the stars from our
sample suggests the star to be not warmer than 4800\,K (on our
adjusted scale). This remains in contrast with what has been found
from photometry.

A similar, albeit less severe case, is HD2796. Our corrected temperature
is about 100\,K cooler, but this star also has a somewhat higher
metallicity ($\mbox{[Fe/H]}\sim-2.5$). If the differences between excitation
temperatures and photometric temperatures was less severe for giants
at higher metallicity, this could explain the discrepant
temperature. However, both our initial and final stellar parameter
solutions as well as that of \citet{cayrel2004} do not match any
position of the red giant branch on the isochrone. Rather, the star sits
slightly above it, perhaps suggesting that the star is descending onto
the horizontal branch. However, \cite{yong13_II} do find a gravity
that that matches the isochrone. Again, the shape of the
Balmer lines suggest a cooler temperature than the photometrically
derived temperatures.

\section{Conclusion}\label{sec:conc}

We have presented a straight-forward method to adjust spectroscopic
temperatures that have been derived from Fe\,I line abundances as a
function of their excitation potential. The main aim was to avoid
obtaining too cool spectroscopic temperatures especially for upper red
giant branch stars that are often several hundred degrees cooler than
photometrically derived temperatures. A pleasant side effect of this
temperature adjustment is that higher surface gravities are obtained
and artificially high microturbulent velocities are avoided. Overall
this brings the fully spectroscopic, reddening independent stellar
parameters in agreement with photometrically derived ones, and also
with isochrones of the respective stellar metallicities. While the
stellar sample is based on stars with $-3.3<\mbox{[Fe/H]}<-2.5$,
additional tests indicate that our adjustment scheme is not usable for
stars with $\mbox{[Fe/H]}<-4.0$.

We tested our method with additional stars from the \citet{cayrel2004}
sample that were also analyzed by \citet{yong13_II} and generally
found agreement to within $\sim$50\,K. The results of applying the new
temperature determination method to new samples of extremely
metal-poor stars will be reported in a forthcoming paper.

\acknowledgements{A.R.C. acknowledges the financial support through
  the Australian Research Council Laureate Fellowship 0992131, and
  from the Australian Prime Minister's Endeavour Award Research
  Fellowship, which has facilitated his research at MIT. This research
  has made use of the SIMBAD database, operated at CDS, Strasbourg,
  France and of NASA's Astrophysics Data System Bibliographic
  Services.}

\textit{Facilities:} \facility{Magellan-Clay (MIKE)}

\end{document}